\PassOptionsToPackage{super,sort&compress,comma}{natbib}
\documentclass[pdflatex,iicol,sn-nature]{sn-jnl}

\usepackage{graphicx}%
\usepackage{multirow}%
\usepackage{amsmath,amssymb,amsfonts}%
\usepackage{amsthm}%
\usepackage{mathrsfs}%
\usepackage[title]{appendix}%
\usepackage{xcolor}%
\usepackage{textcomp}%
\usepackage{manyfoot}%
\usepackage{booktabs}%
\usepackage{algorithm}%
\usepackage{algorithmicx}%
\usepackage{algpseudocode}%
\usepackage{listings}%
\usepackage{newtxtext}  

\theoremstyle{thmstyleone}%
\theoremstyle{thmstyletwo}%

\theoremstyle{thmstylethree}%

\makeatletter
\let\Oldaffil\affil
\renewcommand{\affil}[2][]{\Oldaffil[#1]{{\small #2}}}
\makeatother

\makeatletter
\@ifpackageloaded{geometry}{%
  \geometry{reset, 
    twoside=false, bindingoffset=0pt, centering,
    a4paper, left=25mm, right=25mm, top=25mm, bottom=25mm,
    includehead, includefoot, heightrounded}
}{%
  \usepackage[twoside=false,bindingoffset=0pt,centering,
    a4paper,left=18mm,right=18mm,top=20mm,bottom=20mm,
    includehead,includefoot,heightrounded]{geometry}
}
\makeatother

\begin{document}
\raggedbottom
\title[Stereotyping by strategy standing diversifies cooperation patterns in indirect reciprocity]{Stereotyping by strategy standing diversifies cooperation patterns in indirect reciprocity}







\author[1,4]{\fnm{Ming} \sur{WEI}}

\author*[2,4,5]{\fnm{Xin} \sur{WANG}}\email{wangxin\_1993@buaa.edu.cn}

\author[2,4]{\fnm{Wenqiang} \sur{ZHU}}

\author[2,4,5]{\fnm{Longzhao} \sur{LIU}}

\author[6]{\fnm{Hongwei} \sur{ZHENG}}

\author[9,10,11,12]{\fnm{Feng} \sur{FU}}

\author*[2,3,4,5,7,8]{\fnm{Shaoting} \sur{TANG}}\email{tangshaoting@buaa.edu.cn}

\affil[1]{\orgdiv{School of Mathematical Sciences}, \orgname{Beihang University}, \orgaddress{\city{Beijing}, \postcode{100191}, \country{China}}}

\affil[2]{\orgdiv{School of Artificial Intelligence}, \orgname{Beihang University}, \orgaddress{\city{Beijing}, \postcode{100191}, \country{China}}}

\affil[3]{\orgdiv{Hangzhou International Innovation Institute}, \orgname{Beihang University}, \orgaddress{\city{Hangzhou}, \postcode{311115}, \country{China}}}

\affil[4]{\orgdiv{Key Laboratory of Mathematics, Informatics and Behavioral Semantics}, \orgname{Beihang University}, \orgaddress{\city{Beijing}, \postcode{100191}, \country{China}}}

\affil[5]{\orgdiv{Beijing Advanced Innovation Center for Future Blockchain and Privacy Computing}, \orgname{Beihang University}, \orgaddress{\city{Beijing}, \postcode{100191}, \country{China}}}

\affil[6]{\orgdiv{Beijing Academy of Blockchain and Edge Computing}, \orgaddress{\city{Beijing}, \postcode{100085}, \country{China}}}

\affil[7]{\orgdiv{Institute of Medical Artificial Intelligence}, \orgname{Binzhou Medical University}, \orgaddress{\city{Yantai}, \postcode{264003}, \country{China}}}

\affil[8]{\orgdiv{Institute of Trustworthy Artificial Intelligence}, \orgname{Zhejiang Normal University}, \orgaddress{\city{Hangzhou}, \postcode{310012}, \country{China}}}

\affil[9]{\orgdiv{Department of Mathematics}, \orgname{Dartmouth College}, \orgaddress{\city{Hanover}, \postcode{NH 03755}, \country{USA}}}

\affil[10]{\orgdiv{Department of Biomedical Data Science}, \orgname{Geisel School of Medicine at Dartmouth}, \orgaddress{\city{Lebanon}, \postcode{NH 03756}, \country{USA}}}

\affil[11]{\orgdiv{Department of Applied \& Computational Mathematics, School of Engineering \& Applied Science}, \orgname{Yale University}, \orgaddress{\city{New Haven}, \postcode{CT 06520}, \country{USA}}}

\affil[12]{\orgdiv{Department of Mathematics}, \orgname{Harvard University}, \orgaddress{\city{Cambridge}, \postcode{MA 02138}, \country{USA}}}

\abstract{Indirect reciprocity explains how cooperation evolves through social reputations. People observe others, assign reputations, and condition their future actions on these assignments. This process is cognitively demanding, and stereotyping offers a simpler alternative by replacing individual-level reputation with group-level information. Theoretical models commonly implement stereotyping through exogenously given group labels. In real societies, however, group-level impressions may be associated with observable patterns of behavior. Here we propose a framework of stereotyping by strategy standing, in which mutants may condition their actions on the overall reputation level associated with a resident strategy rather than on the recipient’s reputation. We show that this form of stereotyping can diversify stable cooperation in indirect reciprocity. As the strength of stereotyping increases, additional cooperative evolutionarily stable norm-strategy (ESS) pairs emerge in substantial numbers. In particular, we identify eight highly cooperative ESS pairs that become stable under very weak stereotyping. These pairs, which we call the counterparts of the leading eight, share the same social norms as the classical leading eight and differ only in how they prescribe behavior between bad individuals. They are unstable without stereotyping because they can be invaded by their corresponding leading strategies, but they become stable once stereotyping exceeds a critical threshold. Our results suggest that group-level impressions based on strategy standing can provide a coarse-grained informational route to stable cooperation and offer a more behaviorally grounded perspective on how stereotyping affects indirect reciprocity.}


\onecolumn
\maketitle




\section*{Introduction}\label{intro}
Human societies depend on cooperation among unrelated individuals~\cite{axelrod1981evolution,nowak2006five,rand2013human}. Such cooperation supports collective action, social exchange, and institutional life, but it also creates a persistent evolutionary puzzle. Cooperative behavior can be individually costly while generating benefits for others, so purely self-interested action can undermine socially valuable outcomes~\cite{hamilton1963evolution,fehr2003nature}. One way to reduce this tension is to let individuals condition their behavior on social information beyond the immediate payoff of a single interaction. A central example is reputation~\cite{nowak1998evolution,milinski2002reputation,milinski2016reputation}, which provides information about individuals' past behavior that can guide helping decisions. Stereotyping~\cite{ashmore2015conceptual,fiske1993controlling,judd1993definition,hilton1996stereotypes} offers another form of social information by allowing individuals to use group-level impressions instead of individual-level reputations. These informational mechanisms provide cues for action and can help organize social behavior beyond private interests. 

Reputational information can regulate behavior because individuals may gain future benefits by building a good reputation through prosocial actions~\cite{zhu2024reputation,fu2008reputation}. In evolutionary game theory, indirect reciprocity provides a formal account of how cooperation can evolve through individual reputations~\cite{alexander2017biology,nowak1998evolution,leimar2001evolution,panchanathan2003tale,brandt2004image,brandt2005indirect,ohtsuki2006leading}. In a typical donation game, a donor and a recipient are selected to interact, and the donor decides whether to pay a cost to help the recipient. This decision is made according to the reputational context of the interaction, and the donor’s action is then assessed according to a shared social norm. Depending on the amount of information used in assessment, social norms are commonly classified by the order of information they use. First-order norms, such as image scoring, evaluate only the donor’s action~\cite{nowak1998evolution}. Second-order norms additionally consider the recipient’s reputation, allowing a distinction between unjustified defection and justified refusal to help bad recipients~\cite{leimar2001evolution,panchanathan2003tale}. Third-order norms further take the donor’s own reputation into account, leading to a richer class of assessment rules and to the classical leading eight norms that can sustain cooperation~\cite{ohtsuki2004should,ohtsuki2006leading}. Higher-order norms can therefore make moral assessment more context-sensitive, especially for judging defection, but they may also become more vulnerable to noise and errors~\cite{hilbe2018indirect,fujimoto2023evolutionary,schmid2023quantitative}. This makes accurate information about the recipient’s reputation crucial for the stability of higher-order norms.

Nonetheless, assessing individual reputations can be cognitively demanding. Stereotypes can reduce this burden by allowing people to use group-level impressions as a simplified basis for evaluating others~\cite{tajfel1969cognitive,hamilton1976illusory,fiske1993controlling,macrae1994stereotypes,hilton1996stereotypes,bordalo2016stereotypes,hinton2017implicit}. Since group-level information is usually coarser, stereotypes may distort social judgment and weaken the accurate transmission of behavioral information~\cite{hamilton1976illusory,bordalo2016stereotypes}. At the same time, stereotypes may also serve as cognitively efficient heuristics when individual-level information is costly, noisy, or unavailable~\cite{gigerenzer1996reasoning}. Recent work has brought this idea into indirect reciprocity. By modeling stereotypes as group-level reputations, individuals may approximate the reputations of outgroup members by a single reputation value attached to the group, which can generate ingroup favoritism and intergroup cooperation~\cite{masuda2012ingroup}. Kawakatsu et al. showed that stereotypes can either facilitate or impede cooperation, depending on how reputational information is shared, but that adaptive stereotype use may also spread even when it reduces collective cooperation~\cite{kawakatsu2024stereotypes}. Other models have further explored how prejudicial groups can emerge and coevolve with cooperation under indirect reciprocity~\cite{whitaker2018indirect}. Together, these studies show that group-level impressions can substantially alter the dynamics of reputation-based cooperation. However, the groups that carry such stereotyping impressions are typically introduced as predefined group labels or structured populations. This approach is useful for isolating the effect of replacing individual reputation with group-level information, but it leaves open a complementary question. In reality, stereotyped groups may reflect social roles, repeated observations, or perceived behavioral regularities~\cite{eagly1984gender,jussim2009unbearable}. It therefore remains unclear how cooperation by indirect reciprocity changes when stereotyping is linked not to externally assigned labels, but to group-level impressions that are endogenously generated by behavioral strategies.

\begin{figure*}[t]
\begin{center}
\includegraphics[width = 1\linewidth]{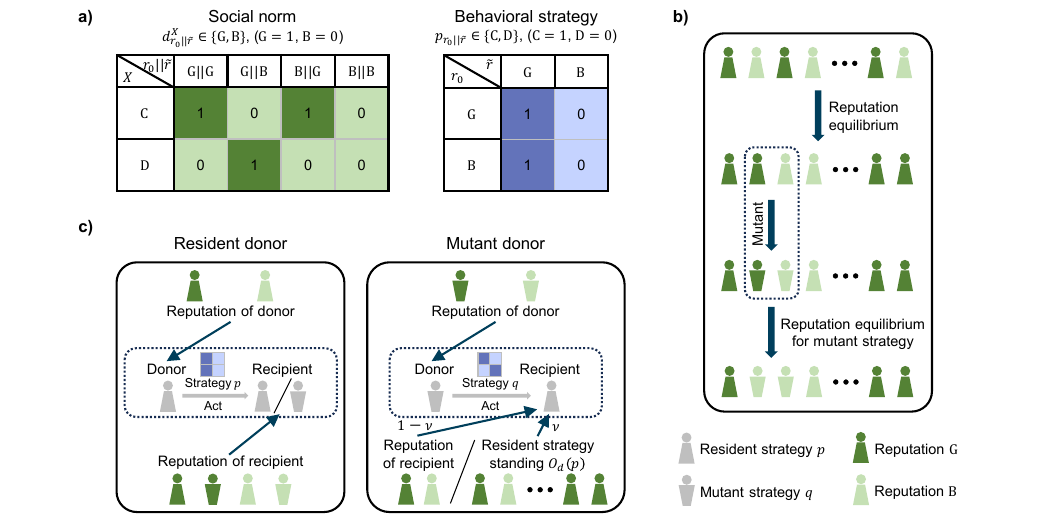}
\end{center}
\caption{\textbf{Indirect reciprocity with stereotyping based on strategy standing.} \textbf{a} Illustration of the third-order framework of social norms and behavioral strategies in indirect reciprocity. In each round, two individuals are randomly selected to play a donation game as donor and recipient. Here, $r_0$ denotes the donor's reputation and $\tilde{r}$ denotes the recipient's reputation. Each can be either good ($\rm G$) or bad ($\rm B$), and together they define the reputational context $r_0||\tilde{r}$ of the interaction. $X$ denotes the donor's action, which can be either cooperation ($\rm C$) or defection ($\rm D$). Each strategy entry $p_{r_0||\tilde{r}}$ specifies the action taken by the donor under the corresponding reputational context, while each norm entry $d_{r_0||\tilde{r}}^X$ specifies the reputation assigned to the action under the corresponding reputational context. \textbf{b}, \textbf{c} Schematics of strategy mutation and the interaction model. We consider an infinitely large population dominated by a resident strategy $p$, into which rare mutants carrying strategy $q$ may appear. The resident population is assumed to have already reached a reputation equilibrium, and mutations do not alter its stability. After a sufficiently long time, rare mutants in the resident environment also reach a reputation equilibrium (\textbf{b}). When a resident acts as donor, it follows the strategy $p$ and conditions its action on the reputational context. Mutants, however, can infer from public information the overall reputation level of the resident population, which we define as the standing of the resident strategy and denote by $O_d(p)$. When a mutant acts as donor, it follows the ordinary reputation-based rule with probability $1-\nu$. With the remaining probability $\nu$, it instead conditions its action on the standing of the current resident strategy. This additional channel represents stereotyping based on strategy standing (\textbf{c}).}
\label{model_schematic}
\end{figure*}

To address this question, we introduce stereotyping by strategy standing into the theoretical framework of third-order indirect reciprocity. We define the standing of a resident strategy as the stationary reputation level reached by a population using that strategy, which is independent of the initial reputation distribution~\cite{ohtsuki2004should}. When a rare mutant acts as donor, it may condition its action either on the recipient’s individual reputation or on the standing of the resident strategy (Fig.~\ref{model_schematic}c). We refer to the probability of using the latter rule as the strength of stereotyping. Under this form of stereotyping, individuals using a resident strategy with high standing are more likely to receive cooperation, regardless of their current individual reputation. This captures a common feature of interactions among strangers: even when people lack direct knowledge of the recipient’s personal reputation, they may still help by relying on a coarse-grained expectation about the surrounding social environment. In our model, strategy standing serves as such a population-level reputational cue. We find that this mechanism expands the set of cooperative evolutionarily stable norm-strategy (ESS) pairs. In particular, eight new ESS pairs appear in the highest performance range in addition to the leading eight. These pairs share exactly the same social norms as the leading eight and differ only in a single entry of their behavioral strategies. We refer to them as the counterparts of the leading eight. Although they are not stable without stereotyping, they can resist invasion by their corresponding leading strategies once the stereotyping strength exceeds a critical threshold. This reveals a counterintuitive role of stereotyping: when group-level impressions are linked to strategy standing, they can expand rather than restrict the space of stable cooperation.

\section*{Model}\label{model}
\subsection*{Interactions and reputations}
We consider an infinitely large, well-mixed population in which individuals engage in one-shot donation games. In each interaction, two players are randomly selected, one as the donor and the other as the recipient. The donor chooses whether to cooperate or defect. Cooperation, denoted by $\rm C$ or $1$, means that the donor pays a cost $c$ to provide a benefit $b$ to the recipient. Defection, denoted by $\rm D$ or $0$, means that the donor does nothing and no benefit is produced. Each individual carries a binary reputation, which can be either good, denoted by $\rm G$, or bad, denoted by $\rm B$. For convenience, we assign numerical values $\rm G=1$ and $\rm B=0$.

The donor’s action is determined by its current behavioral strategy. A strategy $p$ is represented by a table with four entries (Fig.~\ref{model_schematic}a). Each entry $p_{r_0||\tilde{r}}$ takes the value $0$ or $1$ and specifies the action prescribed under a particular reputational context, $r_0||\tilde{r}$. Here, $r_0$ and $\tilde{r}$ denote the current reputations of the donor and the recipient, respectively. 
We allow for execution errors. In each interaction, an intended cooperative action is implemented as defection with probability $\mu_e\ge 0$.

After each interaction, an observer evaluates the donor’s action according to a shared social norm $d$ and updates the donor’s reputation. This information is then made public to the population. The social norm contains eight entries, $d_{r_0||\tilde{r}}^X$ (Fig.~\ref{model_schematic}a). Each entry specifies the reputation assigned to the donor after taking action $X$ under the reputational context $r_0||\tilde{r}$. We also allow for assignment errors. With probability $\mu_a\ge 0$, the observer assigns the wrong reputation, so that an action that should be assessed as bad is instead assigned a good reputation, or vice versa.

\subsection*{Reputation dynamics for single strategy}
Consider a population governed by a social norm $d$ and dominated by a strategy $p$. Let $h_t^p(p)$ denote the fraction of individuals with good reputation at time $t$. Since the population is infinitely large and well-mixed, $h_t^p(p)$ fully characterizes the population-level state of reputation. The temporal change of $h_t^p(p)$ follows
\begin{equation}
    \begin{aligned}
    \frac{\rm d}{{\rm d}t}h_t^p(p) =\ &  h_t^p(p)\Big[h_t^p(p)D_{{\rm G}||{\rm G}}^{p}+(1- h_t^p(p))D_{{\rm G}||{\rm B}}^{p}\Big] + (1-h_t^p(p))\Big[h_t^p(p)D_{{\rm B}||{\rm G}}^{p}+(1- h_t^p(p))D_{{\rm B}||{\rm B}}^{p}\Big]-h_t^p(p) \\
    := &\ I\big(h_t^p(p)\big).
    \end{aligned}
\label{equation:single_err}
\end{equation}
Here, $D_{r_0||\tilde{r}}^p$ denotes the effective social norm entries after incorporating both execution error and assignment error (derivation is given in Materials and methods). The reputation equilibrium of strategy $p$, denoted by $h_*^p(p)$, is obtained by solving $I\big(h_t^p(p)\big)=0$. This equilibrium is unique and independent of the initial reputation state~\cite{ohtsuki2004should}.

\subsection*{Mutation and stereotyping over strategy standing}
In a population dominated by resident strategy $p$, rare mutants carrying a different strategy $q$ may occasionally appear. Following the standard time-scale separation used in previous studies of indirect reciprocity, we assume that reputation dynamics are much faster than strategy change, so that the resident population has already reached its reputation equilibrium $h_*^p(p)$ when mutation occurs~\cite{radzvilavicius2021adherence,wei2025indirect,jiang2026nonlinear,wei2026indirect} (Fig.~\ref{model_schematic}b). Under public reputation, mutants can observe the reputation level of the resident population. We define this population-level reputation as the standing of strategy $p$ under social norm $d$, denoted by $O_d(p)$. 

When a mutant acts as donor, it may either use the recipient’s individual reputation or rely on stereotyping. With probability $\nu$ (stereotyping strength), the mutant does not apply its strategy using the recipient’s individual reputation $\tilde{r}$. Instead, it replaces $\tilde{r}$ with the standing of the resident strategy (Fig.~\ref{model_schematic}c). With probability $O_d(p)$, the donor reads the reputational context as $r_0||{\rm G}$, whereas with probability $1-O_d(p)$, it reads the context as $r_0||{\rm B}$. Note that, when stereotyping is absent ($\nu=0$), this model reduces to the conventional indirect reciprocity dynamics based on individual reputations~\cite{ohtsuki2004should}. When $\nu=1$, mutant donors act entirely according to stereotyping.

Let $h_t^p(q;\nu)$ denote the fraction of mutants with good reputation at time $t$. The reputation dynamics follow 
\begin{equation}
    \begin{aligned}
    \frac{\rm d}{{\rm d}t}h_t^p(q;\nu) =\ & h_t^p(q;\nu)\Big[h_*^p(p)\hat{D}_{\rm G||G}^{q\to p}+(1- h_*^p(p))\hat{D}_{\rm G||B}^{q\to p}\Big] \\
    &+ (1-h_t^p(q;\nu))\Big[h_*^p(p)\hat{D}_{\rm B||G}^{q\to p}+ (1- h_*^p(p))\hat{D}_{\rm B||B}^{q\to p}\Big] - h_t^p(q;\nu), 
    \end{aligned}
\label{equation:mutant_err}
\end{equation}
where $\hat{D}_{r_0||\tilde{r}}^{q\to p}$ denotes the effective social norm entries after incorporating stereotyping, execution error, and assignment error. The stable point of the system~\eqref{equation:mutant_err}, denoted by $h_*^p(q;\nu)$, gives the reputation equilibrium of the mutant strategy in the resident environment (see Materials and methods for detailed definitions and derivation).

\subsection*{ESS analysis}
We denote by $\theta(p,p)$ the probability of cooperation in interactions between resident individuals. Similarly, $\theta(p,q)$ denotes the cooperation probability when a resident donor meets a mutant recipient, and $\theta(q,p)$ denotes the cooperation probability when a mutant donor meets a resident recipient. Definitions of these quantities are given in Materials and methods. The expected payoffs of the resident strategy and the mutant strategy in a single interaction are then given by
\begin{align}
    W(p|p) &= b \theta(p,p)-c \theta(p,p) = (b-c) \theta(p,p), \label{equation:wpp}\\
    W(q|p) &= b \theta(p,q)-c \theta(q,p). \label{equation:wqp}
\end{align}
Strategy $p$ is evolutionarily stable under social norm $d$ if, for every invading strategy $q \neq p$, $W(p|p)>W(q|p)$ holds. In this case, we call $(d,p)$ an evolutionarily stable norm-strategy pair, or an ESS pair.

\begin{figure*}[t]
\begin{center}
\includegraphics[width = 1\linewidth]{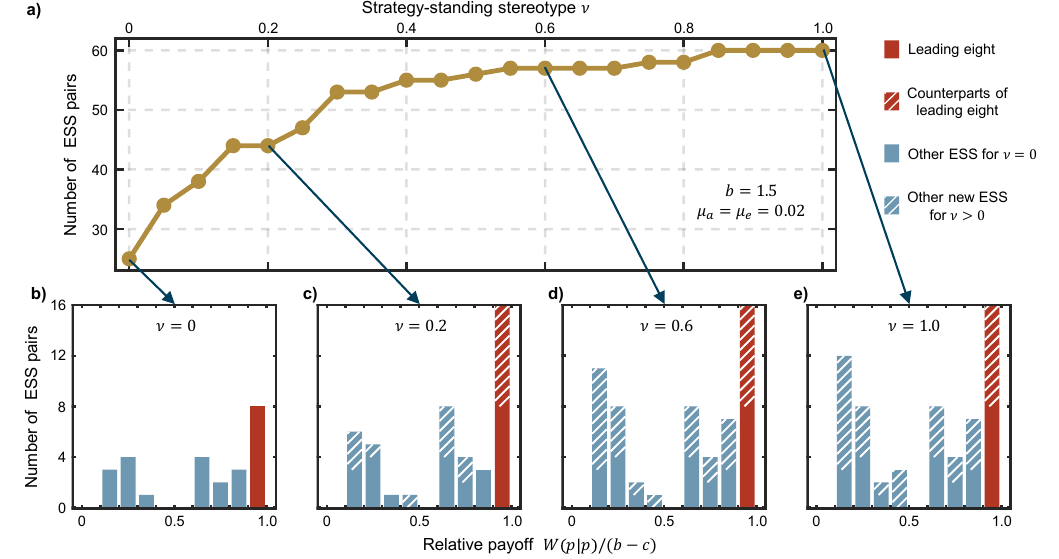}
\end{center}
\caption{\textbf{Stereotyping based on strategy standing stabilizes a broader and more diverse set of highly cooperative ESS pairs.} \textbf{a} Number of cooperative ESS pairs as a function of the stereotyping strength $\nu$. The increase of ESS pairs is non-uniform and exhibits a step-like pattern. As $\nu$ rises from 0 to 0.4, the number of cooperative ESS pairs increases from 25 to 55, corresponding to 30 newly stabilized pairs. By contrast, as $\nu$ further increases from 0.4 to 1, the number rises only from 55 to 60, adding 5 more pairs. \textbf{b}--\textbf{e} Distribution of cooperative ESS pairs across different relative payoff intervals at $\nu=0$ (\textbf{b}), $0.2$ (\textbf{c}), $0.6$ (\textbf{d}), and $1.0$ (\textbf{e}), respectively. Solid bars represent ESS pairs that already exist at $\nu=0$, whereas hatched bars represent ESS pairs that are newly stabilized as $\nu$ increases. A relatively low level of stereotyping is sufficient to stabilize a substantial number of additional cooperative strategies. Initially, the highest payoff interval $[0.9,1.0]$ contains eight ESS pairs, corresponding to the leading eight. When $\nu$ reaches 0.2, another eight ESS pairs appear in the same interval, which we refer to as the counterparts of the leading eight. As $\nu$ increases further, newly stabilized ESS pairs are found only in lower intervals. Parameter values are $b=1.5$, $c=1$, $\mu_a=\mu_e=0.02$.}
\label{ESS_pairs}
\end{figure*}

\section*{Results}\label{results}
\subsection*{Strategy-standing stereotyping expands the set of cooperative ESS pairs}
We first study how the strength of stereotyping affects the evolution of cooperation. For each parameter setting, we perform an exhaustive search for ESS pairs over the full space of social norms and behavioral strategies, following the procedure described in Materials and methods. Note that ALLD strategy (always defect) is evolutionarily stable under any social norm. Since our focus is on the stability of cooperative strategies, we exclude ESS pairs with ALLD from the following analysis and retain only those that sustain a positive level of cooperation.

As the stereotyping strength increases, the number of cooperative ESS pairs also increases, but the trend is non-uniform and step-like (Fig.~\ref{ESS_pairs}a). When $\nu$ is small, many $(d,p)$ pairs that are unstable without stereotyping have already become stable. As $\nu$ increases, the rate at which new ESS pairs are added slows down substantially. We identify $25$, $44$, $57$, and $60$ cooperative ESS pairs, given $\nu=0$, $0.2$, $0.6$, and $1.0$, respectively (Fig.~\ref{ESS_pairs}b--e). This result shows that stereotyping based on strategy standing can stabilize additional cooperative strategies, including a substantial fraction that become stable under weak stereotyping. This qualitative pattern also holds robustly under different parameter settings, including different benefit values and symmetric or asymmetric error rates (Figs.~S1--S3). 

Stereotyping does not remove existing cooperative ESS pairs. Instead, it only adds new stable pairs to the ESS set. Without strategy-standing stereotyping, each social norm can support up to one cooperative ESS pair. However, the same norm can be paired with multiple stable cooperative strategies under stereotyping. Thus, this mechanism does not simply change whether a norm can support cooperation. It also expands the range of behavioral strategies that can remain stable under the same social norm. This distinction is important because the newly stabilized ESS pairs do not emerge by gaining resistance to ALLD. In fact, stereotyping does not improve the performance of cooperative strategies against ALLD, because these free riders do not cooperate regardless of whether group-level information is available. This implies that the new ESS pairs are already able to resist invasion by ALLD even without stereotyping. Their instability at $\nu=0$ is therefore not caused by unconditional defection, but by competition with other cooperative strategies.

\begin{figure*}[t]
\begin{center}
\includegraphics[width = 1\linewidth]{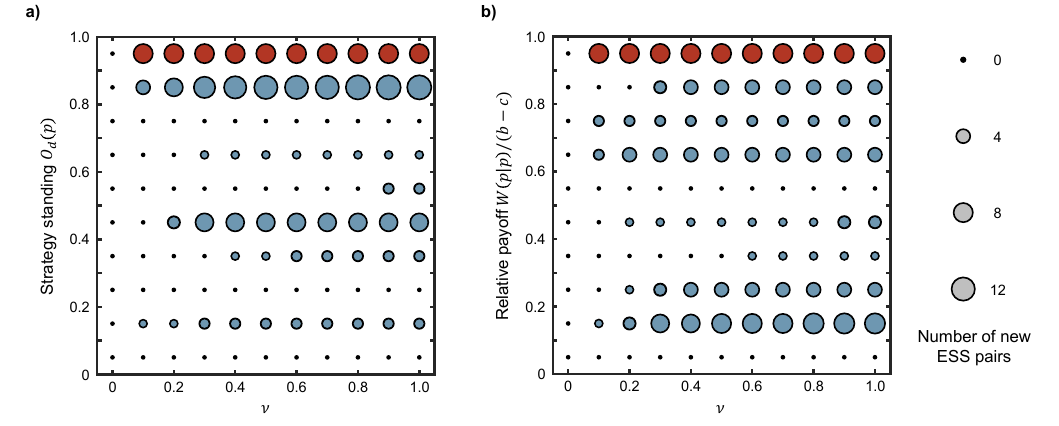}
\end{center}
\caption{\textbf{The counterparts of the leading eight lie in the highest ranges of both strategy standing and relative payoff, and they become stable even at very low stereotyping strength.} The panels show how the number of newly stabilized ESS pairs changes with the stereotyping strength $\nu$, classified by intervals of strategy standing (\textbf{a}) and relative payoff (\textbf{b}). In both panels, the number of newly stabilized ESS pairs in the highest interval, $[0.9,1.0]$, follows the same trend and remains at eight when $\nu\ge 0.1$. They correspond to the same set of ESS pairs, namely the counterparts of the leading eight. Parameters are $b=1.5$, $c=1$, and $\mu_a=\mu_e=0.02$.}
\label{ESS_distr}
\end{figure*}

\subsection*{Counterparts of the leading eight rank highest among new ESS pairs}

Next, we examine the performance of the newly stabilized ESS pairs. We focus on two quantities. One is strategy standing $O_d(p)$, which measures the reputation level reached by a population dominated by a given ESS pair. The other is relative payoff. In a fully cooperative population, every donor pays the cost of cooperation and provides the benefit to the recipient, so the average payoff per interaction reaches its maximum value, $b-c$. We therefore define $W(p|p)/(b-c)$ as the relative payoff, which reflects the level of cooperation in a population dominated by strategy $p$. New ESS pairs gradually emerge as $\nu$ increases, and their relative payoffs are distributed across a wide range between 0 and 1 (Fig.~\ref{ESS_distr}b). When $\nu=1$, 18 of the 35 new ESS pairs have relative payoffs above 0.6, whereas the remaining pairs are below 0.5. Their distribution in strategy standing, however, shows a different pattern. Among these new ESS pairs, 21 reach a strategy standing above 0.8 (Fig.~\ref{ESS_distr}a). This is natural since strategies with high standing are more likely to gain an advantage under stereotyping. Meanwhile, it shows that a good reputational environment does not necessarily imply a high level of cooperation.

\begin{figure*}[t]
\begin{center}
\includegraphics[width = 1\linewidth]{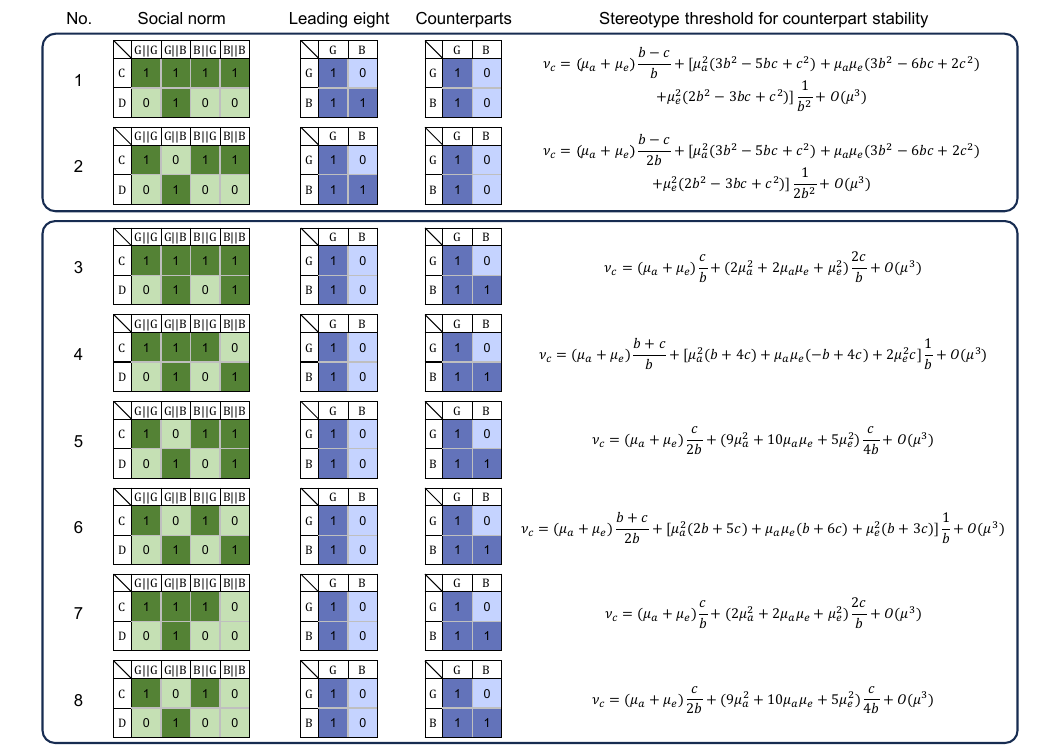}
\end{center}
\caption{\textbf{The counterparts of the leading eight share the same social norms as the leading eight and differ only in a single entry of their behavioral strategies.} The tables show the leading eight ESS pairs and their counterpart pairs. Each counterpart shares the same social norm as its corresponding leading eight pair. Based on the relationship between the leading and counterpart strategies, we divide them into two groups. Within each leading-eight--counterpart correspondence, the two strategies differ only in the action taken when a bad donor encounters a bad recipient ($p_{\rm B||B}$). The other three strategy entries are identical, with $p_{\rm G||G}=p_{\rm B||G}=1$ and $p_{\rm G||B}=0$. Therefore, only two types of strategies are involved, namely the Co-strategy (CO, $p_{\rm B||B}=0$) and the Or-strategy (OR, $p_{\rm B||B}=1$), both of which appear in the leading eight. In group 1, the leading eight pairs adopt OR, whereas their counterparts adopt CO. Conversely, in group 2, the leading eight pairs adopt CO, whereas their counterparts adopt OR. The rightmost column shows the asymptotic expansions of the critical stereotyping strength for each counterpart strategy to resist invasion by its corresponding leading-eight strategy when error rates are small.}
\label{CL_strategy}
\end{figure*}

Nonetheless, eight new ESS pairs form an important exception to this pattern. They lie in the highest intervals of both strategy standing and relative payoff. These pairs become stable even under low stereotyping strength, as shown by the hatched red bars in panels b--e of Fig.~\ref{ESS_pairs} and the red bubbles in Fig.~\ref{ESS_distr}. They also show high cooperation levels and parameter robustness similar to those of the leading eight (hatched red bars in panels b--e of Figs.~S1--S3). This similarity is not accidental. We find that these eight new ESS pairs correspond one-to-one to the leading eight and share exactly the same social norms (Fig.~\ref{CL_strategy}). For convenience, we refer to them as the counterparts of the leading eight, denoted by CL1 to CL8, corresponding to L1 to L8. We also refer to their strategies as counterpart strategies, in contrast to the corresponding leading strategies. Each counterpart strategy differs from its corresponding leading strategy in only one entry, $p_{\rm B||B}$, which specifies the action taken when a bad donor meets a bad recipient. The other three entries are identical and satisfy $p_{\rm G||G}=p_{\rm B||G}=1$ and $p_{\rm G||B}=0$, meaning that individuals cooperate with good recipients and punish bad recipients when they themselves have good reputation. Specifically, the leading strategies L1 and L2 are Or-strategies (OR, $p_{\rm B||B}=1$), whereas their counterparts are Co-strategies (CO, $p_{\rm B||B}=0$). For the remaining cases of L3 to L8, the relationship is reversed: the leading strategies are CO, whereas their counterparts are OR.

In ideal conditions without errors, OR and CO are nearly indistinguishable under the leading eight norms. In populations dominated by these strategies, individuals with bad reputation are so rare that interactions in which a bad donor meets a bad recipient almost never occur. As a result, the value of $p_{\rm B||B}$ has little effect. Once errors are introduced, however, this entry begins to matter. Under the norms corresponding to L3--L8, CO is favored over OR because it avoids the cost of cooperation in the $\rm B||B$ context. By contrast, in scenarios of L1 and L2, cooperation in the $\rm B||B$ context is assessed as good, whereas defection in this context is assessed as bad. The long-term reputational benefit of cooperation can then outweigh the immediate cost, making OR the leading strategy.

The leading eight have been shown to resist invasion by ALLD even under relatively low benefits, while maintaining a cooperation rate close to the maximum level in the resident population~\cite{ohtsuki2004should}. These two properties can be expressed as
\begin{equation}
    \frac{b}{c}>1+\mathcal{O}(\mu), 
\end{equation}
and
\begin{equation}
    W(p|p) = (b-c)[1-\mathcal{O}(\mu)].
\end{equation}
Here, $\mu$ denotes a small error parameter comparable in magnitude to $\mu_a$ and $\mu_e$. The counterparts of the leading eight satisfy the same properties (see Table~S1 for the explicit results), indicating that they have almost the same ability as the leading strategies to resist unconditional defectors and sustain cooperation. This suggests that the main obstacle to the stability of these counterparts is not invasion by ALLD, but competition with their corresponding leading strategies.

\begin{figure*}[t]
\begin{center}
\includegraphics[width = 1\linewidth]{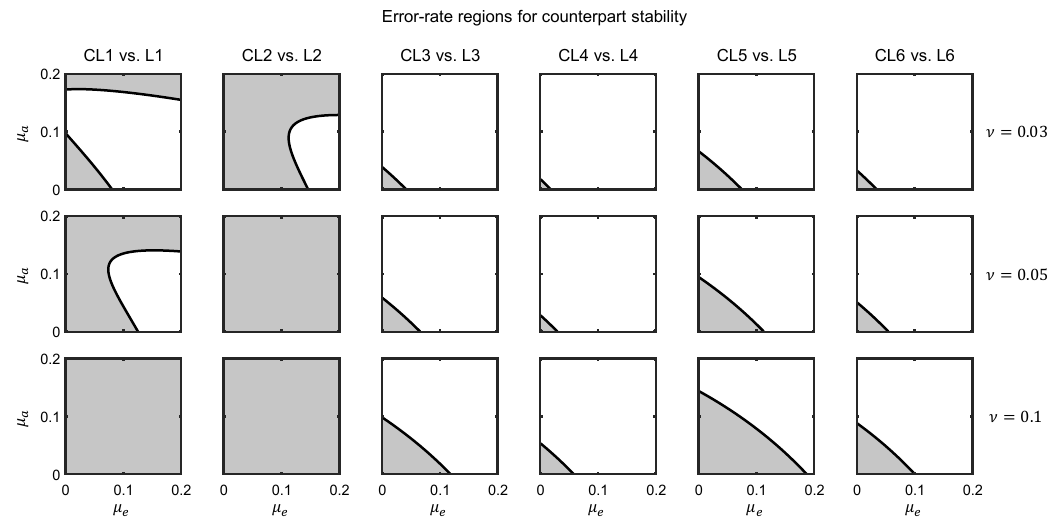}
\end{center}
\caption{\textbf{Strategy-standing stereotyping allows counterpart strategies to remain stable over broader ranges of error rates against their corresponding leading strategies.} Panels show the stability regions (gray) of counterpart strategies under different levels of stereotyping strength. Rows from top to bottom correspond to $\nu=0.03$, $0.05$, and $0.1$. When both error rates are small enough, all counterpart strategies are stable. As $\nu$ increases, they remain stable over increasingly broader regions of the error-rate space. CL1 and CL2 can resist invasion by their leading strategies L1 and L2 over broad error-rate ranges even when the stereotyping strength is very low, and relatively high assignment error further favors the stability of these two counterpart strategies. Results for CL7 and CL8 are not shown because their second-order asymptotic expansions are identical to those of CL3 and CL5, respectively, leading to highly similar stability patterns. Parameter values are $b=1.5$ and $c=1$.}
\label{CL_stable_b1.5}
\end{figure*}

\subsection*{A stereotyping threshold stabilizes the counterpart strategies}
So far, we have shown that stereotyping over strategy standing can make the counterparts of the leading eight evolutionarily stable. This raises a further question: how much stereotyping is required for this stabilization, and how does this requirement depend on errors and payoff parameters? To quantify this requirement, we define the critical stereotyping strength $\nu_c$ as the threshold above which a counterpart strategy becomes stable against its corresponding leading strategy. We first examine this threshold on the error-rate plane by identifying, for each fixed value of $\nu$, the regions in which each counterpart strategy can resist invasion by its corresponding leading strategy. The results show that all counterpart strategies are stable when both execution and assignment errors are sufficiently rare. As $\nu$ increases, these stability regions expand, indicating that more frequent reliance on strategy standing makes it easier for the counterparts to resist invasion (Fig.~\ref{CL_stable_b1.5}). This qualitative pattern is robust to changes in the benefit parameter $b$ (Fig.~S4).

We also find that CL1 and CL2 behave differently from the other counterparts. First, these two strategies show relatively broad stability regions, and their stereotyping thresholds decrease when the assignment error rate becomes sufficiently large (Fig.~S5b). Second, their stability regions shrink as the benefit $b$ increases, whereas the other counterpart strategies show the opposite trend (Fig.~S5a). This contrast originates from the different roles of the strategies CO and OR in these cases. For CL1 and CL2, the counterpart strategy is CO, whereas the corresponding leading strategy is OR. In the $\rm B||B$ context, OR cooperates and CO defects. Under the norms of L1 and L2, cooperation in this context is assessed as good, whereas defection is assessed as bad. Thus, the invading OR strategy can gain a reputational advantage by cooperating in $\rm B||B$, even though it pays the immediate cost of cooperation. A larger benefit $b$ amplifies the future payoff gained from this reputational advantage, making the leading strategy harder to resist and increasing the threshold for CL1 and CL2.

The effect of assignment error adds another layer. When $\mu_a$ is small, increasing $\mu_a$ creates more bad reputations and makes the $\rm B||B$ context more common. This gives the invading OR strategy more opportunities to gain a reputational advantage, which tends to increase the threshold. However, when $\mu_a$ becomes larger, reputations become less reliable. The difference between cooperation and defection in the $\rm B||B$ context is weakened, so the advantage of OR is reduced while its cooperation cost remains. This explains why the thresholds of CL1 and CL2 eventually decline as $\mu_a$ increases.

For the other counterparts, the relation between CO and OR is reversed. The counterpart strategy is OR, whereas the corresponding leading strategy is CO. In these cases, increasing $b$ tends to strengthen the resident counterpart rather than the invader, leading to larger stability regions. Increasing $\mu_a$, by contrast, makes $\rm B||B$ contexts more frequent and exposes OR counterparts more often to the cost of cooperation. This favors the CO leading strategies and therefore raises the stereotyping threshold. 

Theoretical analysis provides further evidence for these qualitative patterns. The exact expression of $\nu_c$ is algebraically lengthy, so we instead present its second-order asymptotic expansion with respect to $\mu_a$, $\mu_e$, $b$, and $c$ (rightmost column in Fig.~\ref{CL_strategy}). In all cases, the first-order term of $\nu_c$ takes the form of $\mu_a+\mu_e$ multiplied by a positive coefficient that depends on $b$ and $c$. This indicates that, when both error rates are small, the critical threshold increases with them, and their effects are comparable. For CL1 and CL2, the first-order coefficients are $(b-c)/b$ and $(b-c)/2b$, respectively, both of which increase with $b$. By contrast, in the remaining cases, the corresponding first-order coefficients decrease as $b$ increases. This difference is consistent with the distinct pattern of CL1 and CL2 observed in numerical results.

\section*{Discussion}\label{disc}

Reputational information plays a central role in sustaining large-scale cooperation in human societies. Indirect reciprocity provides a useful theoretical framework for understanding how altruistic behavior can evolve among unrelated individuals through reputations. However, people may not always evaluate the reputation of every social partner precisely, especially when reputation tracking is cognitively costly or information is incomplete. Instead, they may rely on coarser group-level evaluations, such as group reputation or stereotypes~\cite{masuda2012ingroup,wei2025indirect,katz1933racial,schneider2005psychology,stewart2023group}. Previous theoretical models have represented such group-level information as reputations attached to predefined groups, social categories, or structured populations. This approach captures an important form of cognitive simplification, but it does not fully address the possibility that stereotyped categories may be associated with perceived regularities in behavior, social roles, or past interactions. To examine how such behavior-linked stereotyping affects cooperation by indirect reciprocity, we developed a modeling framework that incorporates stereotyping based on strategy standing. In this framework, individuals may replace a recipient’s individual reputation with the stationary reputation level associated with the resident strategy. This allows the stereotyped cue to be generated endogenously by reputation dynamics, rather than being imposed as an external group label.

Stereotyping can be detrimental to cooperation when it reduces the accuracy of individual reputation. This concern is especially important in real social settings, where stereotypes may reflect biased beliefs, historical inequalities, or categories that individuals cannot control~\cite{eagly1984gender,fiske1993controlling,bordalo2016stereotypes}. Our model does not challenge this concern, but isolates a specific informational mechanism: group-level impressions that are generated by reputation dynamics and linked to behavioral strategies. Within this restricted setting, stereotyping based on strategy standing does not undermine cooperative ESS pairs that are already stable. It further expands the set of stable cooperative strategies. Most notably, this mechanism identifies the counterparts of the leading eight, which achieve cooperation levels and robustness to errors close to those of the classical leading eight. This suggests that the effect of stereotyping on cooperation depends on how group-level impressions are formed and used. When such impressions are linked to behavioral regularities, sacrificing some individual-level precision does not necessarily lead to the collapse of cooperation.

Our findings also have implications for reputation-based cooperation in contemporary social environments. Contemporary social interactions increasingly take place in large-scale settings where individuals cannot directly observe the full behavioral history of others. In online markets, review systems, social media, and platform-mediated communities, aggregated signals are widely used to summarize past behavior at a coarse level~\cite{resnick2000reputation,dellarocas2003digitization}. These signals are not stereotypes in the social-psychological sense, but they share a common information-processing feature. They compress complex individual histories into simplified cues for decision-making. From this perspective, strategy standing can be interpreted as a coarse-grained reputation signal extracted from population-level behavior. Our model suggests that such coarse information does not necessarily weaken cooperation. When group-level cues are tied to behavioral performance, they can help stabilize additional cooperative strategies. At the same time, this result should be interpreted carefully. Stereotyping can be harmful when it is detached from behavior, when it amplifies biased beliefs, or when it assigns individuals to categories that they cannot control. The present framework therefore does not justify stereotyping as a social practice. Rather, it shows that the cooperative effect of group-level information depends strongly on what the group-level signal represents and how it is used in interaction.

In this work, we assumed public reputation, meaning that all individuals share the same assessment of each other. This assumption makes the reputation dynamics tractable and allows us to systematically identify stable norm-strategy pairs. Outside this public setting, reputational information can be private, noisy, or incomplete~\cite{hilbe2018indirect,fujimoto2023evolutionary,wang2023imitation}. The same idea could be extended to such scenarios if individuals maintain a personal repository of reputational assessments. In that case, strategy standing would no longer be a single public value, but an observer-specific summary of how the focal individual evaluates members of a strategy or behavioral type. 
Examining how heterogeneous standings affect cooperation under imperfect information is therefore a promising direction for future work. In addition, our model is based on an infinitely large and well-mixed population. In structured populations, individuals interact locally and reputational information may also spread locally, which could change the effect of stereotyping~\cite{ohtsuki2006simple,su2019spatial,li2020evolution,wang2024evolutionary,wang2026public}. Finally, our model follows the classical indirect reciprocity framework with binary actions and reputations. Extending the analysis to graded helping~\cite{roberts2021benefits}, quantitative reputation~\cite{schmid2023quantitative}, or heterogeneous social environments~\cite{santos2008social} may also lead to richer dynamics.

\section*{Materials and methods}\label{mnm}
In this section, we provide additional details and derivations underlying the Model section.
\subsection*{Reputation dynamics for single strategy}
We first derive the reputation dynamics in a monomorphic population under a fixed social norm $d$. Suppose that all individuals adopt the same behavioral strategy $p$. Let $h_t^p(p)$ denote the fraction of individuals with good reputation at time $t$. 

Consider a short time interval $[t,t+\Delta t]$. During this interval, a fraction $\Delta t$ of individuals are selected as donors and update their reputations after interaction, while the remaining fraction $1-\Delta t$ do not interact and keep their current reputations. If the donor has good reputation, which occurs with probability $h_t^p(p)$, it meets a good recipient with probability $h_t^p(p)$ and a bad recipient with probability $1-h_t^p(p)$. Its prescribed actions are then $p_{{\rm G}||{\rm G}}$ and $p_{{\rm G}||{\rm B}}$, and the resulting assessments are denoted by $d_{{\rm G}||{\rm G}}^{p_{{\rm G}||{\rm G}}}$ and $d_{{\rm G}||{\rm B}}^{p_{{\rm G}||{\rm B}}}$, respectively. Similarly, if the donor has bad reputation, it receives assessments $d_{{\rm B}||{\rm G}}^{p_{{\rm B}||{\rm G}}}$ or $d_{{\rm B}||{\rm B}}^{p_{{\rm B}||{\rm B}}}$ depending on the recipient’s reputation. For compactness, $d_{r_0||\tilde{r}}^{p}$ denotes the assessment assigned to a donor with reputation $r_0$ when it follows the action prescribed by strategy $p$ against a recipient with reputation $\tilde{r}$. Putting these cases together gives 
\begin{equation}
\begin{aligned}
    h_{t+\Delta t}^p(p) = &\Delta t\cdot h_t^p(p)\Big[h_t^p(p)d_{{\rm G}||{\rm G}}^{p}+(1- h_t^p(p))d_{{\rm G}||{\rm B}}^{p}\Big] \\
    &+ \Delta t\cdot(1-h_t^p(p))\Big[h_t^p(p)d_{{\rm B}||{\rm G}}^{p}+(1- h_t^p(p))d_{{\rm B}||{\rm B}}^{p}\Big] + (1-\Delta t)h_t^p(p).
\end{aligned}
\end{equation}
Taking the limit $\Delta t\to 0$, we obtain
\begin{equation}
    \begin{aligned}
    \frac{\rm d}{{\rm d}t}h_t^p(p) = & h_t^p(p)\Big[h_t^p(p)d_{{\rm G}||{\rm G}}^{p}+(1- h_t^p(p))d_{{\rm G}||{\rm B}}^{p}\Big] \\
    &+ (1-h_t^p(p))\Big[h_t^p(p)d_{{\rm B}||{\rm G}}^{p}+(1- h_t^p(p))d_{{\rm B}||{\rm B}}^{p}\Big] - h_t^p(p).
    \end{aligned}
\label{equation:single_noerr}
\end{equation}
We then incorporate two types of errors. Let $\mu_e$ be the execution error rate, so that an intended cooperation is implemented as defection with probability $\mu_e$. Let $\mu_a$ be the assignment error rate, so that the assigned reputation is flipped with probability $\mu_a$. Under these errors, each assessment term $d_{r_0||\tilde{r}}^{p}$ in Eq.~\eqref{equation:single_noerr} is replaced by
\begin{equation}
    D_{r_0||\tilde{r}}^p = (1-2\mu_a)\Big[(1-\mu_e)d_{r_0||\tilde{r}}^p+\mu_e d_{r_0||\tilde{r}}^{\rm D}\Big]+\mu_a.
\end{equation}
The reputation dynamics with errors are therefore Eq.~\eqref{equation:single_err}, in which $I\big(h_t^p(p)\big)$ is a quadratic polynomial in $h_t^p(p)$. The steady-state reputation level of strategy $p$ under norm $d$ is obtained by solving $I\big(h_t^p(p)\big)=0$ and selecting the stable solution. Since $I(0)=D_{{\rm B}||{\rm B}}^{p}\ge \mu_a>0$ and $I(1)=D_{{\rm G}||{\rm G}}^{p}-1\le -\mu_a<0$, a stable solution exists in $h\in(0,1)$. Following the standard analysis of reputation dynamics, this stable equilibrium is unique and independent of the initial reputation state; we denote it by $h_*^p(p)$~\cite{ohtsuki2004should}. This quantity gives the long-run fraction of good reputations in a population fully occupied by strategy $p$.

\subsection*{Mutation and stereotyping over strategy standing}
We next consider the reputation dynamics of a rare mutant strategy $q$ in a population dominated by resident strategy $p$. The resident population is assumed to have already reached its reputation equilibrium $h_*^p(p)$. We define the standing of the resident strategy under norm $d$ as $O_d(p)=h_*^p(p)$. Because mutants are rare, the reputation distribution of residents is not affected by their presence. Thus, a mutant donor faces a resident recipient with good reputation with probability $h_*^p(p)$, and with bad reputation with probability $1-h_*^p(p)$. 

Stereotyping is introduced only when a mutant donor interacts with a resident recipient. With probability $1-\nu$, the mutant donor follows its strategy $q$ using the recipient’s individual reputation $\tilde{r}$. With probability $\nu$, it instead uses the standing $O_d(p)$ of the resident strategy. The effective probability that a mutant donor cooperates is therefore
\begin{equation}
\rho_{r_0||\tilde{r}}^{q\to p}=(1-\nu)q_{r_0||\tilde{r}}+\nu\Big[O_d(p)q_{r_0||{\rm G}}+\big(1-O_d(p)\big)q_{r_0||{\rm B}}\Big].
\label{equation:rho_qp}
\end{equation}
This expression states that, under stereotyping, the mutant donor responds to the resident strategy’s standing rather than to the recipient’s individual reputation. When $\nu=0$, Eq.~\eqref{equation:rho_qp} reduces to the conventional reputation-based rule. When $\nu=1$, the mutant donor acts entirely according to strategy standing.

Given $\rho_{r_0||\tilde{r}}^{q\to p}$, the probability that a mutant donor with reputation $r_0$ receives a good reputation after meeting a resident recipient with reputation $\tilde{r}$ is
\begin{equation}
\hat{d}_{r_0||\tilde{r}}^{q\to p}=\rho_{r_0||\tilde{r}}^{q\to p}d_{r_0||\tilde{r}}^{\rm C}+\big(1-\rho_{r_0||\tilde{r}}^{q\to p}\big)d_{r_0||\tilde{r}}^{\rm D}.
\end{equation}
Including execution error and assignment error gives
\begin{equation}
\hat{D}_{r_0||\tilde{r}}^{q\to p}
=
(1-2\mu_a)
\Big[
(1-\mu_e)\hat{d}_{r_0||\tilde{r}}^{q\to p}
+
\mu_e d_{r_0||\tilde{r}}^{\rm D}
\Big]
+\mu_a.
\label{equation:tilde_D}
\end{equation}

Let $h_t^p(q;\nu)$ denote the fraction of mutants with good reputation at time $t$. Their reputation dynamics are given by Eq.~\eqref{equation:mutant_err}. For compactness, define
\begin{align}
    T_1^p(q) &= h_*^p(p)\hat{D}_{\rm G||G}^{q\to p}+(1- h_*^p(p))\hat{D}_{\rm G||B}^{q\to p}, \\
    T_2^p(q) &= h_*^p(p)\hat{D}_{\rm B||G}^{q\to p}+(1- h_*^p(p))\hat{D}_{\rm B||B}^{q\to p}.
\end{align}
Then Eq.~\eqref{equation:mutant_err} becomes
\begin{equation}
\frac{{\rm d}}{{\rm d}t}h_t^p(q;\nu)=-\big(1-T_1^p(q)+T_2^p(q)\big)h_t^p(q;\nu)+T_2^p(q).
\end{equation}
Thus, once the resident equilibrium $h_*^p(p)$ is known, the stationary reputation level of the mutant strategy can be obtained explicitly as
\begin{equation}
h_*^p(q;\nu)=\frac{T_2^p(q)}{1-T_1^p(q)+T_2^p(q)}.
\end{equation}
This quantity gives the long-run fraction of good reputations among mutants in the resident population.

\subsection*{ESS analysis}
The reputation equilibria derived above can help evaluate whether a resident strategy $p$ is evolutionarily stable under a fixed social norm $d$. We first compute the cooperation rates in the resident-mutant configuration. Here, cooperation rates refer to realized cooperation probabilities after execution error. In a monomorphic resident population, the cooperation rate between resident individuals is
\begin{equation}
    \theta(p,p) = (1-\mu_e)\Big[p_{\rm G||G}(h_*^p(p))^2+(p_{\rm G||B}+p_{\rm B||G}) h_*^p(p)(1-h_*^p(p))+p_{\rm B||B}(1-h_*^p(p))^2\Big].
\end{equation}
When a resident donor meets a mutant recipient, the resident still follows its original strategy $p$. Thus,
\begin{equation}
\begin{aligned}
    \theta(p,q) =\ &(1-\mu_e)\Big[p_{\rm G||G}h_*^p(p)h_*^p(q;\nu)+p_{\rm G||B}h_*^p(p)(1-h_*^p(q;\nu)) \\
    &+ p_{\rm B||G}(1-h_*^p(p))h_*^p(q;\nu) + p_{\rm B||B}(1-h_*^p(p))(1-h_*^p(q;\nu))\Big].
\end{aligned}
\end{equation}
When a mutant donor meets a resident recipient, its action rule is affected by stereotyping over the resident strategy standing. The cooperation rate is therefore
\begin{equation}
\begin{aligned}
\theta(q,p) = &(1-\mu_e)\Big[\rho_{\rm G||G}^{q\to p}h_*^p(q;\nu)h_*^p(p)
+\rho_{\rm G||B}^{q\to p}h_*^p(q;\nu)\big(1-h_*^p(p)\big) \\
&+\rho_{\rm B||G}^{q\to p}\big(1-h_*^p(q;\nu)\big)h_*^p(p)
+\rho_{\rm B||B}^{q\to p}\big(1-h_*^p(q;\nu)\big)\big(1-h_*^p(p)\big)\Big],
\end{aligned}
\end{equation}
where $\rho_{r_0||\tilde{r}}^{q\to p}$ is defined in Eq.~\eqref{equation:rho_qp}. Note that stereotyping is applied only to mutant donors. Nevertheless, it can also affect $\theta(p,q)$ indirectly through the mutant reputation equilibrium $h_*^p(q;\nu)$.

Given these cooperation rates, the expected payoff of the resident strategy in the resident population is given by Eq.~\eqref{equation:wpp}, and the expected payoff of the mutant strategy $q$ in the resident population is given by Eq.~\eqref{equation:wqp}. We say that strategy $p$ is evolutionarily stable under social norm $d$ if, for every mutant strategy $q\neq p$, $W(p|p)>W(q|p)$ holds. In this case, the pair $(d,p)$ is called an evolutionarily stable norm-strategy pair, or an ESS pair.

For any given set of parameter values, we perform an exhaustive search by examining all possible norm-strategy pairs $(d,p)$ according to the ESS criterion above. All pairs that satisfy this condition are recorded as ESS pairs. Note that these results contain mirror-symmetric ESS pairs (see Appendix A in Supporting information for definition)~\cite{ohtsuki2004should}. To better reveal the structure of the ESS set, we prune mirror symmetry in all reported results.

\subsection*{Supporting information}
Appendix A, Figs.~S1--S5, and Table~S1. 

\subsection*{Data availability}
There are no empirical data associated with this study. All results are obtained from computational simulations based on the methods described.

\subsection*{Code availability}
The MATLAB and Maple code used to generate the computational and symbolic results in this study is available through GitHub at \url{https://github.com/RoyWey1998/strategy-standing-stereotyping.git}.

\subsection*{Author contributions}
M.W. and X.W. conceived the initial idea for this study. M.W. developed the model and carried out the theoretical analysis. M.W. wrote the original draft of the paper. M.W., X.W., W.Z., and F.F. revised and edited the paper. X.W., L.L., H.Z., and S.T. provided funding support.
 
\subsection*{Additional information}
\noindent \textbf{Acknowledgements} This work is supported by National Science and Technology Major Project (2022ZD0116800), Program of National Natural Science Foundation of China (12425114, 12301305, 62441617, 12501702), the Fundamental Research Funds for the Central Universities, Beijing Natural Science Foundation (Z230001), the Opening Project of the State Key Laboratory of General Artificial Intelligence (Project No. SKLAGI2025OP16), and Beijing Advanced Innovation Center for Future Blockchain and Privacy Computing. 
\medskip

\noindent \textbf{Competing interests} The authors declare no competing interests.
\medskip

\noindent \textbf{Correspondence and requests for materials} should be addressed to X.W. or S.T.

\bibliography{bib1}

\end{document}


\setstretch{1.2}

\title[Stereotyping by strategy standing diversifies cooperation patterns in indirect reciprocity]{\textbf{\qquad \quad \quad \;\ Supporting Information for}\vspace{2mm}
\\
Stereotyping by strategy standing diversifies cooperation patterns in indirect reciprocity}







\author[1,4]{\fnm{Ming} \sur{WEI}}

\author*[2,4,5]{\fnm{Xin} \sur{WANG}}\email{wangxin\_1993@buaa.edu.cn}

\author[2,4]{\fnm{Wenqiang} \sur{ZHU}}

\author[2,4,5]{\fnm{Longzhao} \sur{LIU}}

\author[6]{\fnm{Hongwei} \sur{ZHENG}}

\author[9,10,11,12]{\fnm{Feng} \sur{FU}}

\author*[2,3,4,5,7,8]{\fnm{Shaoting} \sur{TANG}}\email{tangshaoting@buaa.edu.cn}

\affil[1]{\orgdiv{School of Mathematical Sciences}, \orgname{Beihang University}, \orgaddress{\city{Beijing}, \postcode{100191}, \country{China}}}

\affil[2]{\orgdiv{School of Artificial Intelligence}, \orgname{Beihang University}, \orgaddress{\city{Beijing}, \postcode{100191}, \country{China}}}

\affil[3]{\orgdiv{Hangzhou International Innovation Institute}, \orgname{Beihang University}, \orgaddress{\city{Hangzhou}, \postcode{311115}, \country{China}}}

\affil[4]{\orgdiv{Key Laboratory of Mathematics, Informatics and Behavioral Semantics}, \orgname{Beihang University}, \orgaddress{\city{Beijing}, \postcode{100191}, \country{China}}}

\affil[5]{\orgdiv{Beijing Advanced Innovation Center for Future Blockchain and Privacy Computing}, \orgname{Beihang University}, \orgaddress{\city{Beijing}, \postcode{100191}, \country{China}}}

\affil[6]{\orgdiv{Beijing Academy of Blockchain and Edge Computing}, \orgaddress{\city{Beijing}, \postcode{100085}, \country{China}}}

\affil[7]{\orgdiv{Institute of Medical Artificial Intelligence}, \orgname{Binzhou Medical University}, \orgaddress{\city{Yantai}, \postcode{264003}, \country{China}}}

\affil[8]{\orgdiv{Institute of Trustworthy Artificial Intelligence}, \orgname{Zhejiang Normal University}, \orgaddress{\city{Hangzhou}, \postcode{310012}, \country{China}}}

\affil[9]{\orgdiv{Department of Mathematics}, \orgname{Dartmouth College}, \orgaddress{\city{Hanover}, \postcode{NH 03755}, \country{USA}}}

\affil[10]{\orgdiv{Department of Biomedical Data Science}, \orgname{Geisel School of Medicine at Dartmouth}, \orgaddress{\city{Lebanon}, \postcode{NH 03756}, \country{USA}}}

\affil[11]{\orgdiv{Department of Applied \& Computational Mathematics, School of Engineering \& Applied Science}, \orgname{Yale University}, \orgaddress{\city{New Haven}, \postcode{CT 06520}, \country{USA}}}

\affil[12]{\orgdiv{Department of Mathematics}, \orgname{Harvard University}, \orgaddress{\city{Cambridge}, \postcode{MA 02138}, \country{USA}}}

\maketitle

\clearpage
\setcounter{tocdepth}{2}

\section*{Appendix A. Mirror symmetry}
In the full space considered in this study, a social norm $d$ contains eight binary entries and a behavioral strategy $p$ contains four binary entries. Therefore, there are $2^8\times 2^4=4096$ possible norm-strategy pairs $(d,p)$. We use $1$ to represent good reputation (${\rm G}$) and $0$ to represent bad reputation (${\rm B}$). This numerical representation is only a convention. If the labels $1$ and $0$ are completely exchanged, the resulting norm-strategy pair is a mirror image of the original one~\cite{ohtsuki2004should}.

For a given pair $(d,p)$, its mirror image $(d',p')$ is defined by
\begin{align}
    d_{{\rm G}||{\rm G}}^{\prime X} &= 1-d_{{\rm B}||{\rm B}}^X, \\
    d_{{\rm G}||{\rm B}}^{\prime X} &= 1-d_{{\rm B}||{\rm G}}^X, \\
    d_{{\rm B}||{\rm G}}^{\prime X} &= 1-d_{{\rm G}||{\rm B}}^X, \\
    d_{{\rm B}||{\rm B}}^{\prime X} &= 1-d_{{\rm G}||{\rm G}}^X,
\end{align}
and
\begin{align}
    p'_{{\rm G}||{\rm G}} &= p_{{\rm B}||{\rm B}}, \\
    p'_{{\rm G}||{\rm B}} &= p_{{\rm B}||{\rm G}}, \\
    p'_{{\rm B}||{\rm G}} &= p_{{\rm G}||{\rm B}}, \\
    p'_{{\rm B}||{\rm B}} &= p_{{\rm G}||{\rm G}}.
\end{align}
Here, $X\in\{{\rm C},{\rm D}\}$.

Equivalently, $(d',p')$ is obtained by replacing every occurrence of good reputation by bad reputation, and every occurrence of bad reputation by good reputation, while also reversing the reputation assigned by the social norm. Two mirror-symmetric pairs have the same reputation dynamics up to this relabeling of reputation states. Thus, if $(d,p)$ leads to a trajectory $h_t$, its mirror image $(d',p')$ leads to the corresponding trajectory $1-h_t$ under the exchanged interpretation of good and bad reputations. As a result, the two pairs are dynamically equivalent and have the same evolutionary stability properties. 

In the reported results, we prune mirror-symmetric duplicates and keep only one representative from each mirror-symmetric class.

\clearpage
\begin{figure}[htbp]
\begin{center}
\includegraphics[width = 1\linewidth]{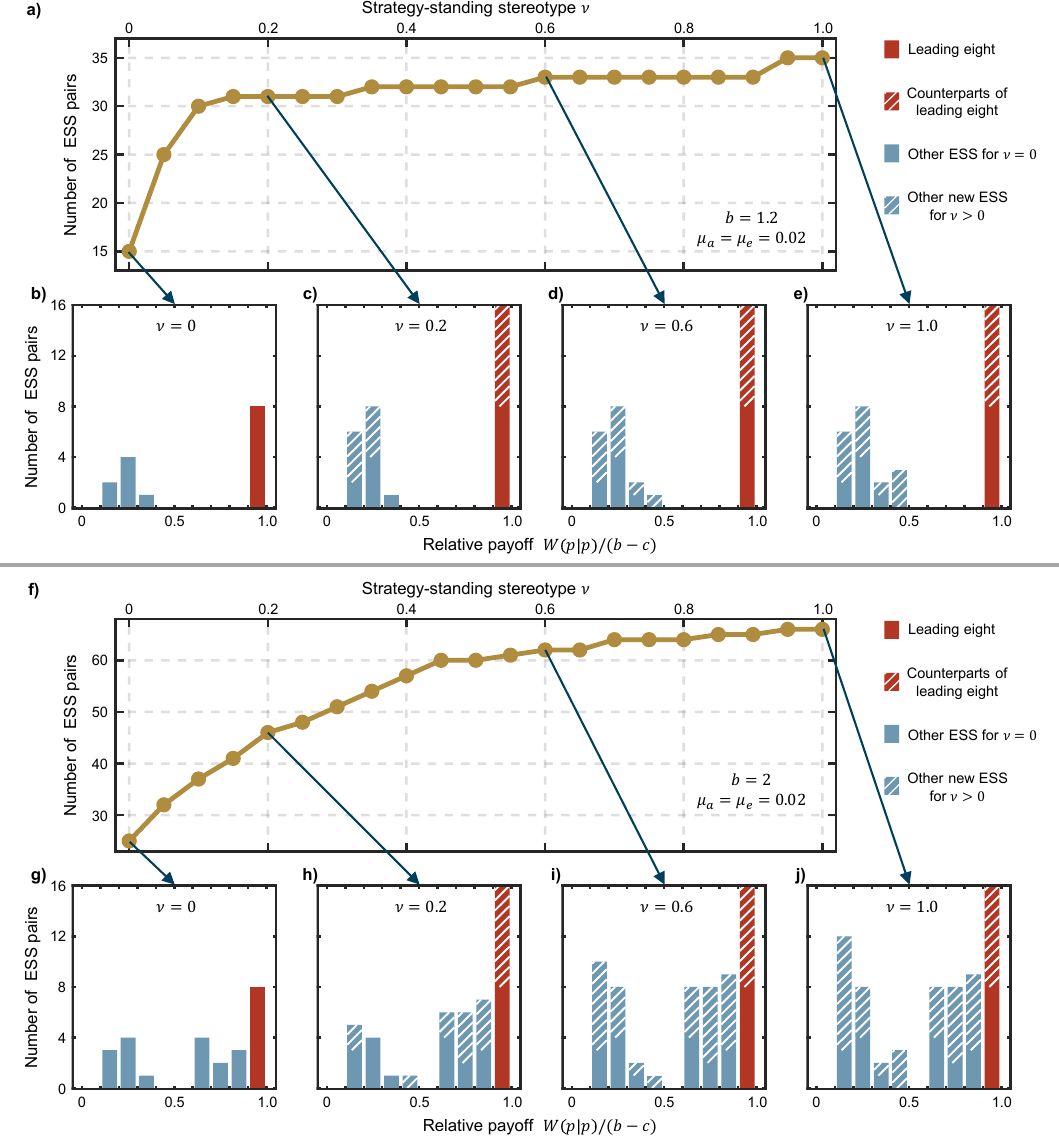}
\end{center}
\caption{\textbf{Evolutionary pattern remains qualitatively robust across different benefit values.} We show the evolutionary outcomes under different values of the benefit parameter. For $b=1.2$ (\textbf{a}--\textbf{e}) and $b=2$ (\textbf{f}--\textbf{j}), the number of cooperative ESS pairs continues to increase with the stereotyping strength $\nu$, although the growth rate gradually slows down as $\nu$ becomes larger. This trend is more pronounced for smaller values of $b$. From $\nu=0$ to $\nu=1$, the number of cooperative ESS pairs increases by 20 for $b=1.2$ and by 41 for $b=2$. In both cases, the counterparts of the leading eight are unstable at $\nu=0$, but become stable once the stereotyping strength increases slightly. These patterns are qualitatively consistent with the result under the baseline parameter setting in the main text. Parameter values: $c=1$, $\mu_a=\mu_e=0.02$.}
\label{sfig1}
\end{figure}

\clearpage
\begin{figure}[htbp]
\begin{center}
\includegraphics[width = 1\linewidth]{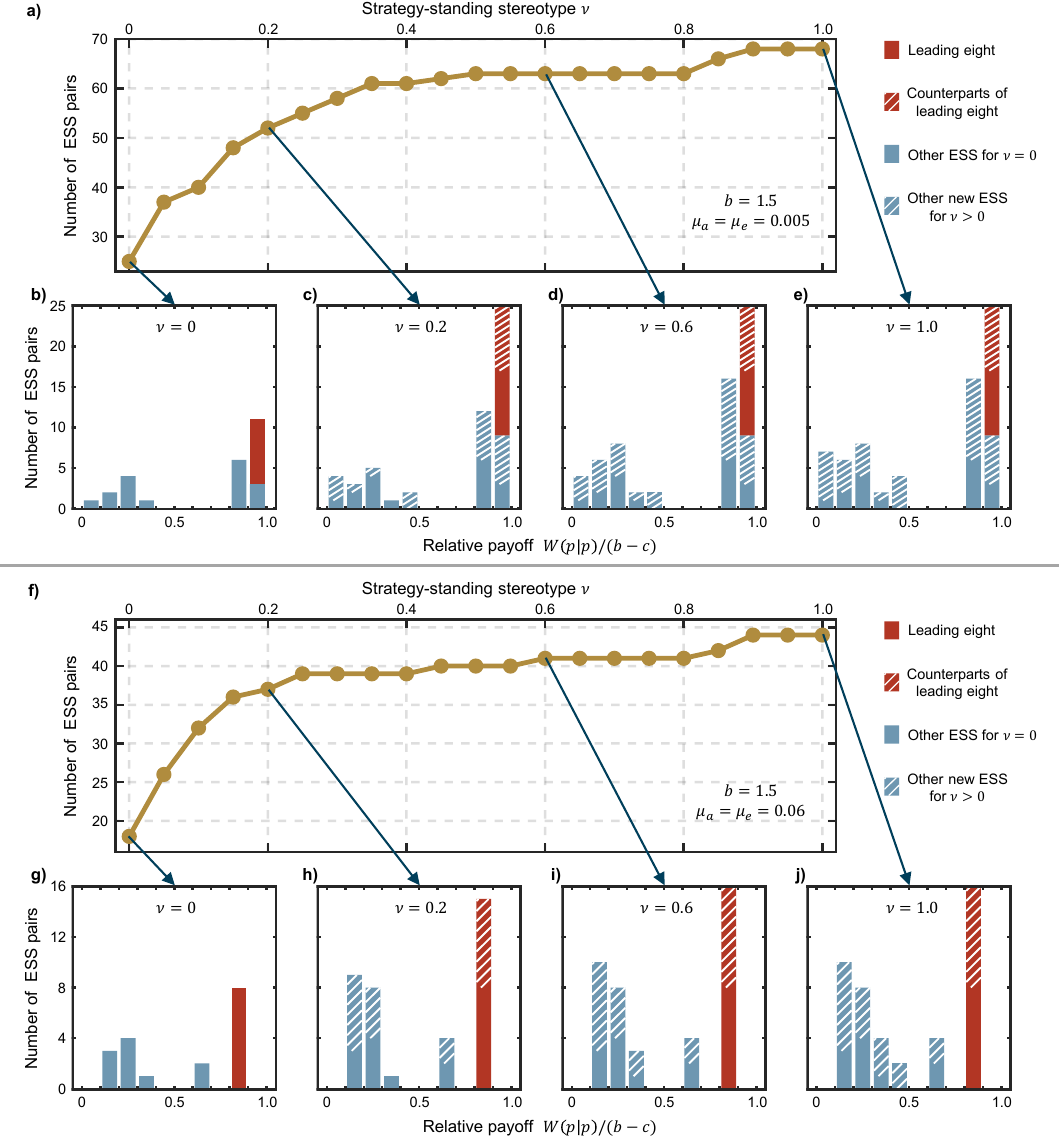}
\end{center}
\caption{\textbf{Evolutionary pattern remains qualitatively robust across different levels of symmetric error rates.} We show the evolutionary outcomes under different symmetric error rates ($\mu_a=\mu_e$). The overall dependence of cooperative ESS pairs on the stereotyping strength $\nu$, as well as the stability of the counterparts of the leading eight, remains qualitatively consistent with the baseline result. When the error rate is smaller ($\mu_a=\mu_e=0.005$, \textbf{a}--\textbf{e}), the number of cooperative ESS pairs is markedly larger, and more additional ESS pairs reach the highest relative-payoff interval. By contrast, when the error rate is larger ($\mu_a=\mu_e=0.06$, \textbf{f}--\textbf{j}), the number of cooperative ESS pairs is reduced, and the relative payoffs of both the leading eight and their counterparts decrease. Nevertheless, they still exhibit the highest relative payoffs. Parameter values: $b=1.5$, $c=1$.}
\label{sfig2}
\end{figure}

\clearpage
\begin{figure}[htbp]
\begin{center}
\includegraphics[width = 1\linewidth]{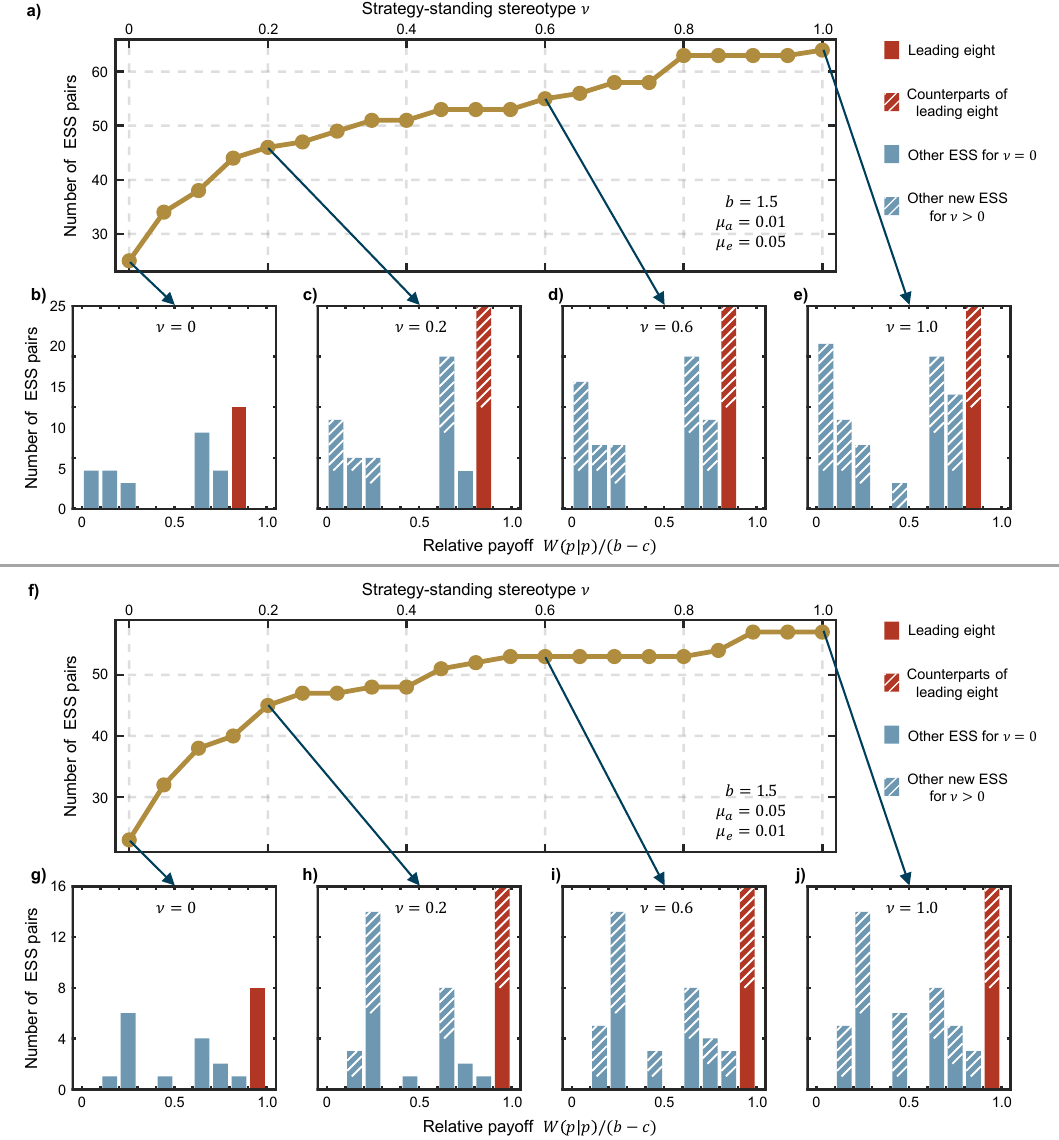}
\end{center}
\caption{\textbf{Evolutionary pattern remains qualitatively robust under asymmetric error rates.} We show the evolutionary outcomes under different combinations of asymmetric error rates ($\mu_a\neq\mu_e$). The error rates are set to $\mu_a=0.01$, $\mu_e=0.05$ for \textbf{a}--\textbf{e} and $\mu_a=0.05$, $\mu_e=0.01$ for \textbf{f}--\textbf{j}. The overall dependence of cooperative ESS pairs on the stereotyping strength $\nu$, as well as the stability of the counterparts of the leading eight, remains qualitatively consistent with the baseline result. An increase in the execution error rate has a stronger effect on relative payoff than an increase in the assessment error rate. For $\mu_a=0.01$ and $\mu_e=0.05$, the relative payoffs of both the leading eight and their counterparts decrease to the interval $[0.8,0.9)$. Nevertheless, they still remain the best-performing ESS pairs in the population. Parameter values: $b=1.5$, $c=1$.}
\label{sfig3}
\end{figure}

\clearpage
\begin{figure}[htbp]
\begin{center}
\includegraphics[width = 1\linewidth]{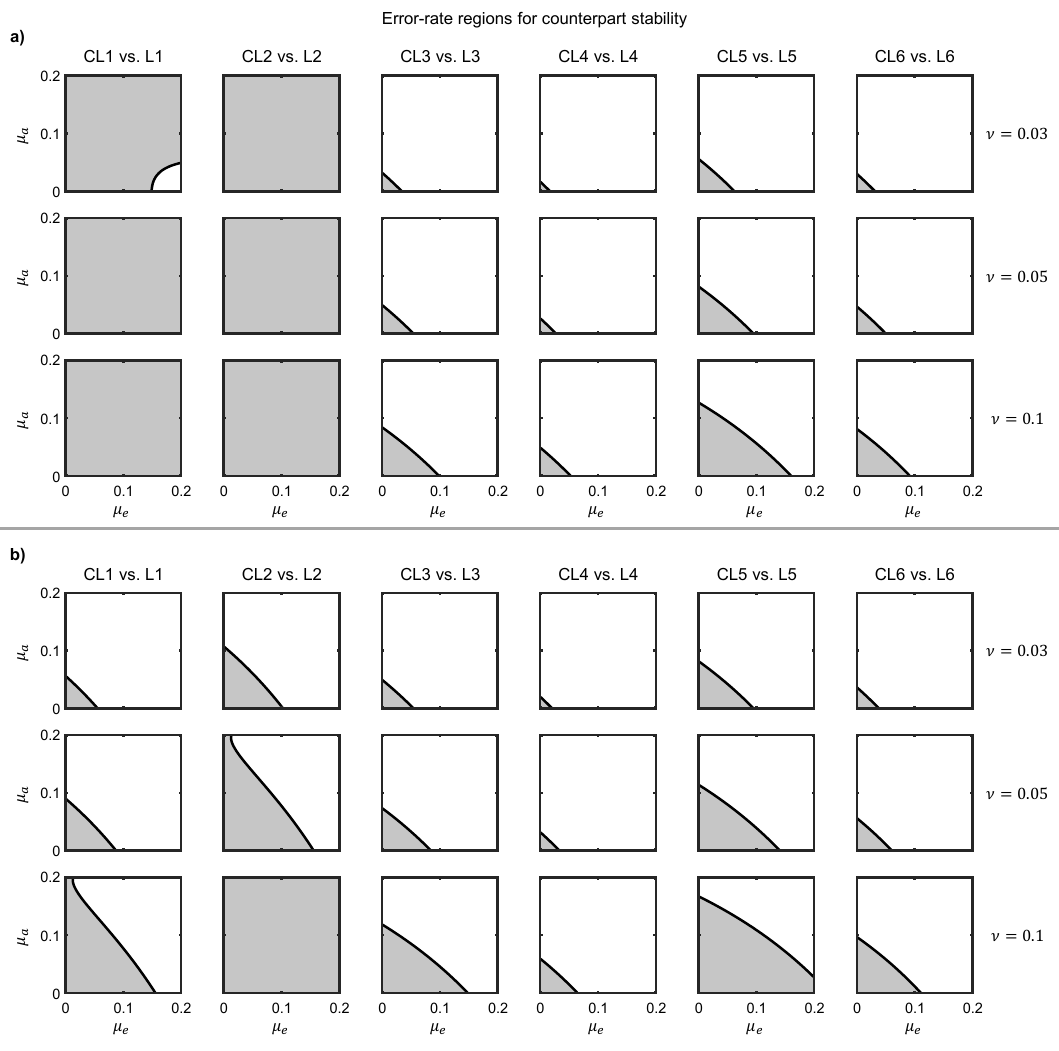}
\end{center}
\caption{\textbf{The benefit parameter has opposite effects on the stability of CO and OR counterpart strategies.} Panels show the stability regions of counterpart strategies under different levels of stereotyping strength, with gray regions indicating the error rate ranges in which a counterpart strategy can resist invasion by its corresponding leading strategy. The benefit parameter is set to $b=1.2$ in \textbf{a} and $b=2$ in \textbf{b}. Within both parts, rows from top to bottom correspond to $\nu=0.03$, $0.05$, and $0.1$, respectively, and the stability regions expand as $\nu$ increases. When $b$ is smaller, the CO counterparts CL1 and CL2 remain stable over broader error-rate ranges. By contrast, when $b$ is larger, the OR counterparts CL3--CL6 show broader stability regions. Results for CL7 and CL8 are not shown because their stability patterns are highly similar to those of CL3 and CL5, respectively. The cost parameter is set to $c=1$.}
\label{sfig4}
\end{figure}

\clearpage
\begin{figure}[htbp]
\begin{center}
\includegraphics[width = 1\linewidth]{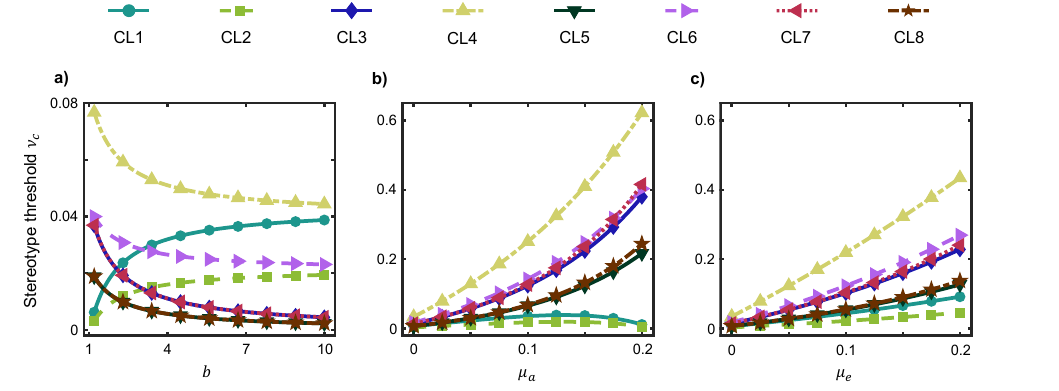}
\end{center}
\caption{\textbf{Robustness of stereotyping strength threshold to benefit and error parameters.} \textbf{a}--\textbf{c} Critical value of stereotyping strength required for each counterpart strategy to resist invasion by its corresponding leading strategy, shown as a function of benefit $b$ (\textbf{a}), assessment error rate $\mu_a$ (\textbf{b}), and execution error rate $\mu_e$ (\textbf{c}). With respect to changes in $b$ and $\mu_a$, CL1 and CL2 show trends that differ clearly from those of the other counterpart strategies. Other parameters: $b=1.5$, $c=1$, $\mu_a=0.02$, and $\mu_e=0.02$.}
\label{sfig5}
\end{figure}

\begin{table}[htbp]
\centering
\caption{\textbf{Conditions for resisting ALLD and expected payoff of the counterparts of the leading eight.} Here, $\mu$ denotes a small error parameter of the same order as $\mu_a$ and $\mu_e$.}
\label{tab:CL_cooperation}
\begin{tabular}{p{0.04\textwidth}|p{0.44\textwidth}|p{0.44\textwidth}}
\hline
Name & Condition for resisting ALLD & Expected payoff $W(p|p)$ \\
\hline
CL1 & $\frac{b}{c}>1+(\mu_a)+(3\mu_a^2+2\mu_a\mu_e)+\mathcal{O}(\mu^3)$ & $(b-c)[1-(\mu_a+2\mu_e)+(-\mu_a^2+2\mu_a\mu_e+\mu_e^2)+\mathcal{O}(\mu^3)]$ \\
CL2 & $\frac{b}{c}>1+(\mu_a)+(3\mu_a^2+2\mu_a\mu_e)+\mathcal{O}(\mu^3)$ & $(b-c)[1-(\mu_a+2\mu_e)+(-\mu_a^2+2\mu_a\mu_e+\mu_e^2)+\mathcal{O}(\mu^3)]$ \\
CL3 & $\frac{b}{c}>1+(3\mu_a+2\mu_e)+(8\mu_a^2+6\mu_a\mu_e+3\mu_e^2)+\mathcal{O}(\mu^3)$ & $(b-c)[1-(\mu_a+2\mu_e)+(\mu_a^2+6\mu_a\mu_e+3\mu_e^2)+\mathcal{O}(\mu^3)]$ \\
CL4 & $\frac{b}{c}>1+(3\mu_a+2\mu_e)+(10\mu_a^2+10\mu_a\mu_e+5\mu_e^2)+\mathcal{O}(\mu^3)$ & $(b-c)[1-(\mu_a+2\mu_e)+(4\mu_a\mu_e+2\mu_e^2)+\mathcal{O}(\mu^3)]$ \\
CL5 & $\frac{b}{c}>1+(3\mu_a+2\mu_e)+(8\mu_a^2+6\mu_a\mu_e+3\mu_e^2)+\mathcal{O}(\mu^3)$ & $(b-c)[1-(\mu_a+2\mu_e)+(\mu_a^2+6\mu_a\mu_e+3\mu_e^2)+\mathcal{O}(\mu^3)]$ \\
CL6 & $\frac{b}{c}>1+(3\mu_a+2\mu_e)+(10\mu_a^2+10\mu_a\mu_e+5\mu_e^2)+\mathcal{O}(\mu^3)$ & $(b-c)[1-(\mu_a+2\mu_e)+(4\mu_a\mu_e+2\mu_e^2)+\mathcal{O}(\mu^3)]$ \\
CL7 & $\frac{b}{c}>1+(2\mu_a+\mu_e)+(7\mu_a^2+6\mu_a\mu_e+2\mu_e^2)+\mathcal{O}(\mu^3)$ & $(b-c)[1-(\mu_a+2\mu_e)+(4\mu_a\mu_e+2\mu_e^2)+\mathcal{O}(\mu^3)]$ \\
CL8 & $\frac{b}{c}>1+(2\mu_a+\mu_e)+(7\mu_a^2+6\mu_a\mu_e+2\mu_e^2)+\mathcal{O}(\mu^3)$ & $(b-c)[1-(\mu_a+2\mu_e)+(4\mu_a\mu_e+2\mu_e^2)+\mathcal{O}(\mu^3)]$ \\
\hline
\end{tabular}
\end{table}


\addcontentsline{toc}{section}{References}
\bibliography{bib2}